\documentclass[12pt,aasmacros]{spieman}  % 12pt font required by SPIE;
\usepackage{amsmath,amsfonts,amssymb}
\usepackage{graphicx}
\usepackage{setspace}
\usepackage{tocloft}
\usepackage{xcolor}
\usepackage{xspace}
\usepackage{lineno}
\usepackage{multirow}
\title{Tuning of Nuclear Spectroscopic Telescope Array (NuSTAR) Application Specific Integrated Circuits (ASICs) to improve low energy threshold of future Hard X-ray Imaging Detectors}

\author[1]{Daniel P. Violette$^{\star}$\textsuperscript{$\dagger$}}
\author[1]{Branden Allen}
\author[1]{Jaesub Hong}
\author[2]{Hiromasa Miyasaka}
\author[1]{Jonathan Grindlay}

\affil[1]{Center for Astrophysics $|$ Harvard \& Smithsonian, 60 Garden St, Cambridge, MA 02138, USA}
\affil[2]{California Institute of Technology, Pasadena, CA 91125, USA}

\cftpagenumbersoff{figure}
\cftpagenumbersoff{table} 
\begin{document} 
\maketitle

\begin{abstract}

Detector commanding, processing and readout of spaceborne instrumentation is often accomplished with Application Specific Integrated Circuits (ASICs). The ASIC designed for the Nuclear Spectroscopic Telescope Array (\textit{NuSTAR}) mission (NuASIC) enables future tiled CdZnTe (CZT) detector array readout for x-ray detectors such as the High Resolution Energetic X-ray Imager (HREXI). Modified NuASIC gain settings have been implemented for HREXI's broader targeted imaging energy range (3-300 keV) compared to \textit{NuSTAR} (2-79 keV), which may require updated NuASIC internal parameters for optimal energy resolution. To reach HREXI’s targeted low energy threshold, we have also enabled the NuASIC’s ``Charge Pump Mode'' (CPM),  which introduces an additional tuning parameter. In this paper, we describe the mechanics of the NuASIC's adjustable parameters and use our recently developed ASIC Test Stand (ATS) to probe a ``bare'' NuASIC using its internal test pulser. We record the effects of parameter tuning on the device's electronics noise and low energy threshold and report the optimal set of parameters for HREXI's updated gain setting. We detail a semi-automated procedure to derive the optimal parameters for each of HREXI's large area, closely tiled NuASIC/CZT detectors to expedite instrument integration.

\end{abstract}

%SPIE

{\noindent \footnotesize{$^{\star}$daniel.violette@cfa.harvard.edu}} \\
{\noindent \footnotesize{\textsuperscript{$^\dagger$}NASA FINESST Fellow}}

\section{Introduction \& Motivation}

Time-domain and multi-messenger astrophysics are  priority science objectives for astronomy over the next decade\cite{Astro2020} and will illuminate the most extreme astrophysical processes. To contribute to this discovery regime, we are developing a large area closely tiled CdZnTe (CZT) imaging detector, the High Resolution Energetic X-ray Imager (HREXI). HREXI is the detector for a  wide-field coded aperture telescope, the HREXI SmallSat Extremes Explorer (\textit{HSEE}), currently in formulation (Grindlay et al.\,2023)\cite{Grindlay23}. The \textit{HSEE} telescope enables broad-band (3-300 keV) imaging with high spatial and spectral resolution to enable pioneering new studies of long and short Gamma-Ray Bursts (GRB), blazar flares, outbursts from magnetars, super-luminal supernovae, tidal disruption events and low-level x-ray flares from black hole low  mass x-ray binaries to measure their poorly understood population and formation. The broad-band imaging of \textit{HSEE} will extend detections and sharpen locations of short x-ray/gamma-ray bursts coincident with gravitational wave detections from merging neutron star (or neutron star--black hole) binaries. Broad-band spectra and $\leq$40 arcsec locations of high redshift long GRBs will allow us to trace the epoch of formation of the very first (Pop III) stars\cite{Bromm2002}.

\begin{figure*}[tbh!]
    \centering
    \includegraphics[width=6.5in]{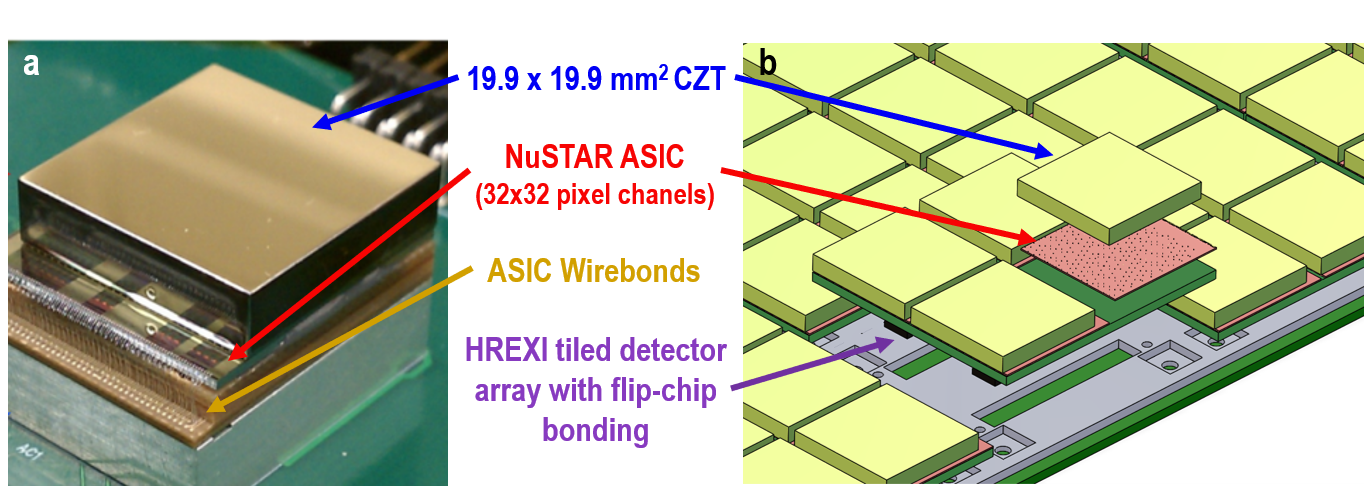}
\caption{(\textbf{\textit{a}}) A single detector crystal unit of a CZT crystal bonded with conductive epoxy to a NuASIC. The NuASIC is connected with 87 wirebonds to a detector PCB to provide ASIC power, command and control, and readout of the detector. (\textbf{\textit{b}}) Computer Aided Design model of the HREXI coded aperture CZT/NuASIC detector plane, composed of tiled arrays of 2\,$\times$\,2 Detector Crystal Arrays (DCA). For HREXI, each NuASIC will have the option of either using 87 Through Silicon Vias (TSVs) or wirebonds. TSV-NuASICs for HREXI enables flip-chip bonding each DCA to the PCB in tightly packed detector spacing.\cite{Violette22}}
    \label{f:detector}
\end{figure*}

\textit{HSEE} draws on technology studied for large area CZT arrays for the \textit{EXIST} telescope proposed for Astro2010\cite{Grindlay11} and initially developed with the \textit{ProtoEXIST} \textit{1} \& \textit{2} balloon-borne x-ray telescope experiments\cite{Hong09,Allen10,Allen11,Hong13,Hong17}. The \textit{HSEE} detector plane will be composed of a closely tiled array of 19.9~$\times$~19.9~mm$^{2}$, 3~mm thick pixelated CZT detectors, each read out by an Application Specific Integrated Circuit (ASIC) originally developed\cite{Cook98,Cook99,Harrison10} for the Nuclear Spectroscopic Telescope Array (\textit{NuSTAR}) mission\cite{Harrison13}. The electrical connection of each of the individual CZT pixels (0.55~$\times$~0.55~mm$^{2}$ pixel active area) to the system for readout is completed with conductive-epoxy bonds to gold studs attached to a matching pattern of input pads (32~$\times$~32 pads, 0.6~mm pitch) on the upper surface of the \textit{NuSTAR} ASIC (NuASIC). The full \textit{HSEE} detector plane will consist of a 16~$\times$~16 detector array (960 cm$^{2}$) compatible with the effective area and sensitivity requirements of a coded aperture telescope. An example of a single CZT detector produced for the \textit{ProtoEXIST2} detector plane is shown in Figure \ref{f:detector}a and a model of the tiled array of such detectors to realize the \textit{HSEE} detector plane is given in Figure \ref{f:detector}b.

The operational parameters of the NuASIC (e.g., threshold, gain select, etc.) are controlled with a single user-programmable command register that enables the adjustment of 15 separate parameters, as well as an enable bit for each of the 1024 pixels and control of the internal test-pulse routing.  The optimization of a NuASIC's resolution and low energy threshold is accomplished in this work through adjustment of a set of these on-board parameters. While optimal settings were determined manually for \textit{NuSTAR's} two telescopes, each with 2~$\times$~2 CZT/ASIC detectors, \textit{HSEE's} larger 256 detector array will require systematic and automated parameter selection for each CZT/ASIC for time- and cost-saving during instrument integration. 

\textit{HSEE's} science objectives require sensitivity over a broader range of energies (3-300 keV) than \textit{NuSTAR} (2-79 keV). NuASICs flown on \textit{HSEE} will operate with slightly reduced signal gains in order to achieve this expanded dynamic energy range; 4 different global preamplifier gain settings are selectable through a 2-bit gain select register ({\tt gainsel}). Fundamentally, the {\tt gainsel} setting modifies the feedback capacitance value of the NuASIC pixel's charge sensitive preamplifier. As the feedback capacitance is increased the preamplifier output current is conversely reduced. This allows for a broader range of photon energies before saturating the NuASIC's 12-bit analog-to-digital converter (ADC). To achieve the desired dynamic pixel energy range of up to $\sim$350 keV, the \textit{HSEE} detectors will operate with a {\tt gainsel} setting of 1, compared to the highest gain setting used on \textit{NuSTAR}, {\tt gainsel} = 0. In this paper, we evaluate how our modified NuASIC gain setting will affect the optimal set of NuASIC internal parameters and explore device optimization to achieve the best resolution and low energy threshold. 

We explore the sensitivity of the NuASIC to these internal parameter adjustments in a normal operating configuration as compared to a ``Charge Pump Mode'' (CPM) that further reduces the standing current at the preamplifier output which, in turn, suppresses detector noise and improves energy resolution. For this study, we have tested a Through-Silicon Via (TSV) enabled NuASIC (described by Hong et al.\,2021)\cite{Hong21} that has been modified to have the 87 commanding, power, and readout lines routed through the NuASIC's silicon substrate to its back surface through a via-last process\cite{Ovental21}. This TSV-enabled design will ultimately improve instrument integration by replacing the previously required 87 individual wire bonds per detector with a simpler and more robust flip-chip bonding process. To probe a ``bare'' unattached TSV-NuASIC we have designed and built an ASIC Test Stand (ATS) for control and data readout that contacts the device using an array of micro-pogo probes\cite{Violette18,Violette22}. We utilize the NuASIC's internal test pulser to simulate fixed-energy x-ray events in each pixel channel and to determine the impact of NuASIC parameter adjustments.

We next provide an overview of the operation of the NuASIC in order to clarify how x-ray events are captured and recorded. This includes a description of the five tuning parameters that may be adjusted within the NuASIC to aid in stabilizing operation and achieving optimal energy resolution and low energy threshold. We then provide a summary of the results obtained from operating the NuASIC in normal mode and CPM through pixel channel analog output and the NuASIC's digital readout. Lastly, we discuss the effect of choosing the optimal parameter set for the NuASIC in both normal mode and CPM, and how these different operational modes will impact the detector's resolution and low energy threshold.

\section{NuASIC Parameter Overview}

\begin{figure*}[tbh!]
    \centering
    \includegraphics[width=6.5in]{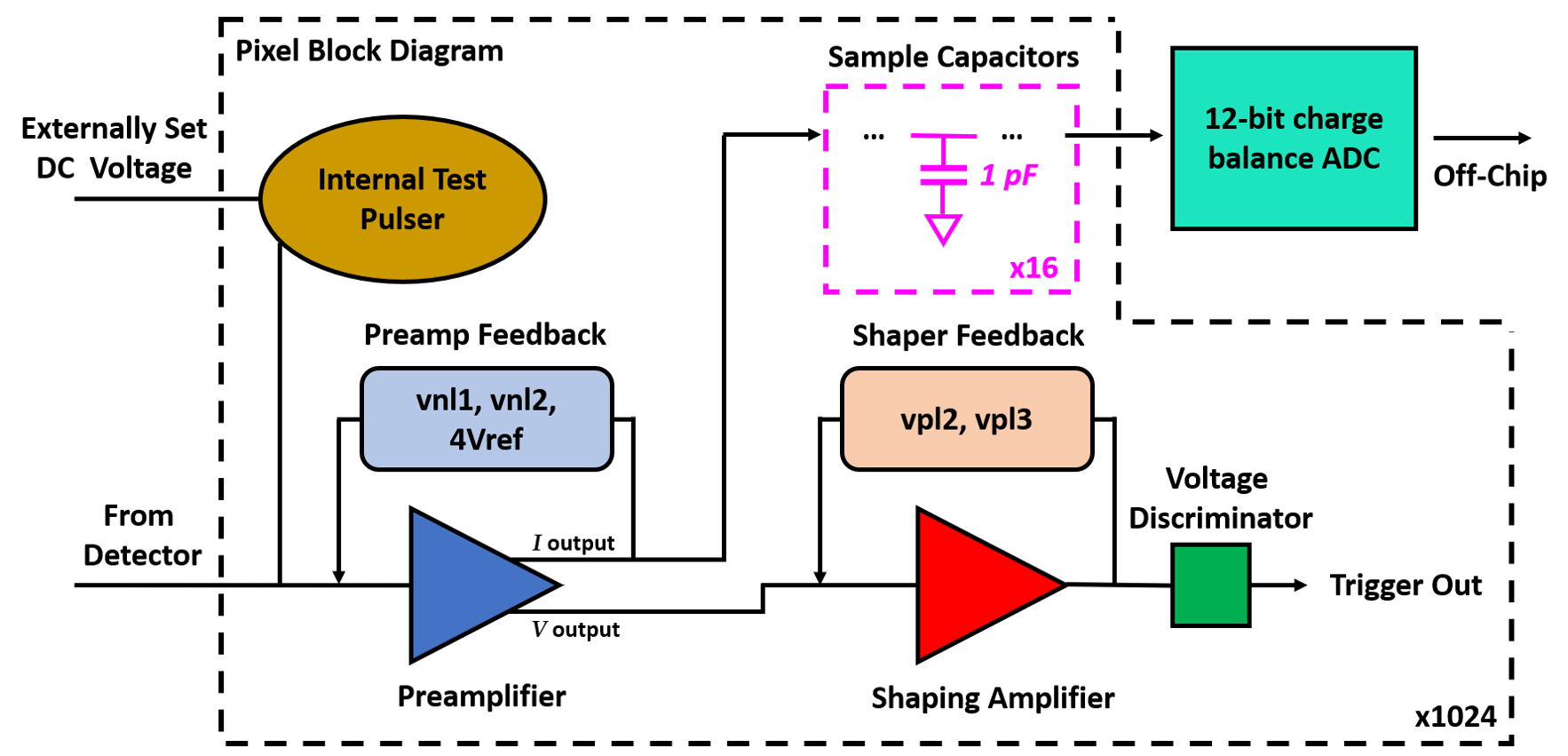}
    \caption{A simplified diagram of the circuity for a single NuASIC pixel showing the feedback parameters. Charge from an incoming event or generated by an internal test pulser is amplified and integrated with a charge-sensitive preamplifier. The preamplifier produces both a voltage and current output signal.  The voltage signal is routed to a shaping amplifier which in turn is attached to a comparator for single-pixel event-detection.  Simultaneously the preamplifier current output signal sets a charge level sequentially on a bank of 16-capacitors.  The sampling is halted after the detection of an event, preserving a series of baseline measurement of the preamplifer baseline charge level as well as recording the integrated charge collected during the event. These capacitors are sampled with a 12-bit charge-rebalanced ADC, after which the output is sent off-chip and the capacitors are reset to the baseline voltage level. The preamplifier standing current is adjusted by the {\tt vnl1sel} (vnl1 bias), and {\tt vnl2sel} (vnl2 bias) command registers. Additionally, the {\tt 4Vsel} register is tuned in CPM to minimize voltage ramping on the preamplifier output by reducing DC bias fluctuations at the input. The shaping amplifier pulse duration and linearity can be modified by adjusting the vpl2 and vpl3 biases ({\tt vpl2sel} and {\tt vpl3sel}, respectively). These parameters are defined in Table \ref{table:1} and apply globally to all 1024 pixels.}
    \label{f:asicdiagram}
\end{figure*}

To understand the tunable parameters of the NuASIC, we first explain the device's basic operation. When any of the NuASIC's 1024 pixel channels receives charge from a photon event or an internally driven test pulser, a series of steps allows for the capture of the signal which can be seen in Figure \ref{f:asicdiagram}. The event charge is received by the pixel preamplifier and produces a voltage and current output. The voltage output is routed through a shaping amplifier to a  voltage discriminator. If the signal surpasses the required programmable threshold, a trigger will be sent off-chip to an external controller (a Complex Programmable Logic Device, CPLD) to begin the NuASIC's event processing. Before an event is triggered, baseline charge from the preamplifier is sequentially collected on a bank of 16 capacitors with a 1 $\mu$s cadence driven by a sampling clock. When a triggering signal arrives, the sampling clock runs for eight additional cycles to capture the event's charge before stopping. When the capacitor bank's charge is then sampled through the NuASIC's 12-bit charge-balance ADC, it contains eight pre-event baseline charge samples and eight charge samples from the event before being returned to a baseline voltage level. The captured baseline and event data are then transmitted off-chip where the baseline values are subtracted from the signal values to determine the energy of the event.  A duplicate current output is returned to the preamplifier input in a feedback loop, which has its resistance set by a smaller value resistor, a current divider sequence, and an amplifier\cite{Cook98}. This feed back loop balances out pixel leakage current and generates the standing current baseline, which is the DC current at the amplifier input or output when no signal from the attach detector or internal pulser is received. After a $\sim$1.8~ms post-trigger processing delay (for a single event reading out a 3~$\times$~3 pixel patch), the NuASIC then returns to an "armed" state until triggered by the next x-ray event. For a more detailed description of the fundamentals underlying the NuASIC design and operation, see Cook et al.\,1998.\cite{Cook98}

\begin{figure*}[tbh!]
   \centering
   \includegraphics[width=6.5in]{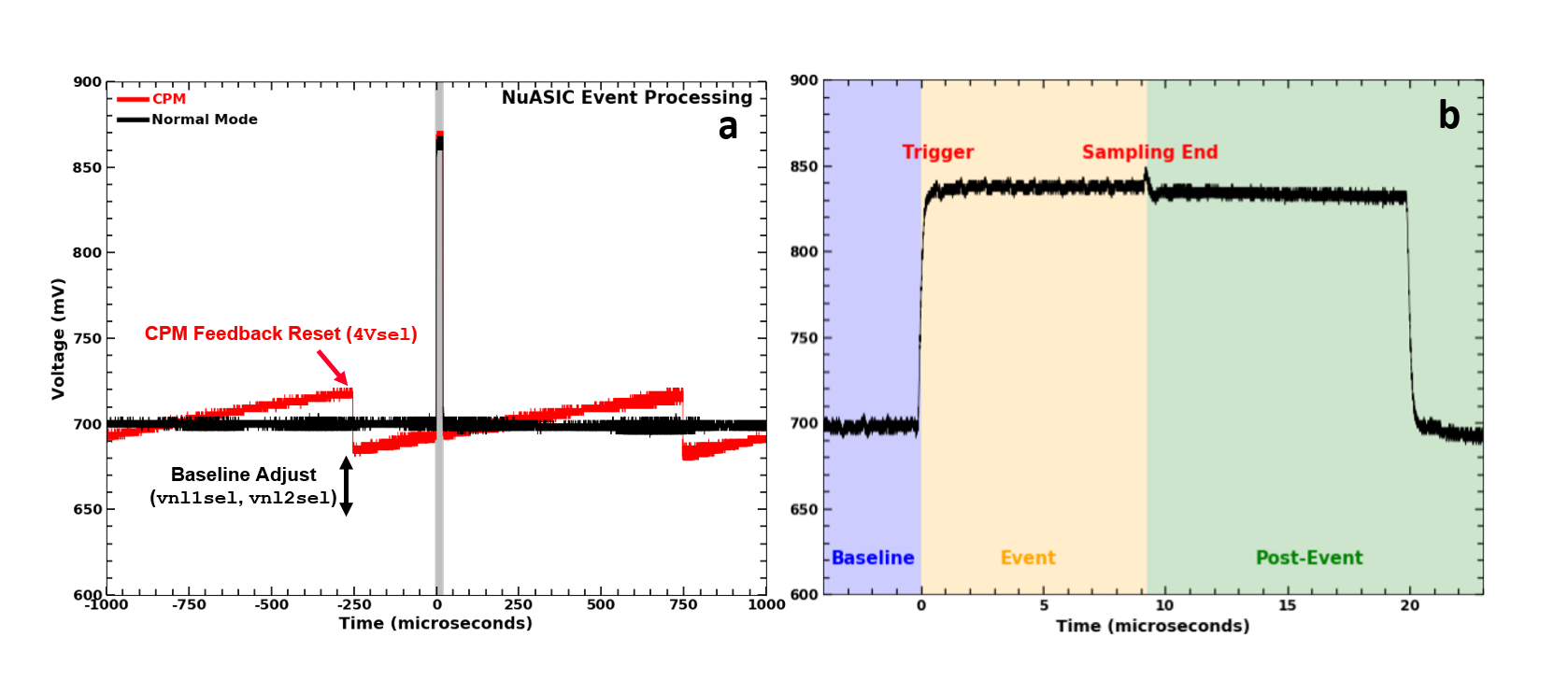}
\caption{(\textbf{\textit{a}}) Analog preamplifier readout of an example event trigger for the NuASIC, containing the event trigger in the center (gray region). Both examples of normal mode and CPM operation are included with the voltage ramping effect of CPM visible (red trace). The baseline voltage level can be adjusted with the {\tt vnl1sel} and {\tt vnl2sel} NuASIC settings, while CPM voltage ramping perturbations can be minimized with {\tt V4sel}. (\textbf{\textit{b}}) Zoomed-in view of the normal mode event trigger. The baseline samples, event samples, and the post-event trace are respectively shown in the blue, yellow, and green shaded regions. Sampling of the pixel channel capacitor bank ends when the sampling clock stops eight clock cycles after the rising signal edge.}
    \label{f:eventexample}
\end{figure*}

The NuASIC supports a single output line which may be commanded to output the preamplifier and shaping amplifier signals for any of the individual pixels. We use this output as an additional diagnostic tool which enables us to directly observe the pixel-by-pixel changes to these outputs which can be compared to shifts in the sampling capacitor's baseline voltage level though the NuASIC's 12-bitcharge-rebalance ADC readout as a function of our parameter adjustments. Figure \ref{f:eventexample}a shows an example of the oscilloscope-captured traces of this preamplifier analog output readout with \ref{f:eventexample}b providing a zoomed-in view of the process. With the exception of the individual pixel enable and test pulser enable masks, the adjustable parameters of the NuASIC are global, and applied simultaneously to all pixels. The standing current at the preamplifier input is increased by the vnl1 bias current, which is used to balance pixel leakage current, and the vnl2 bias current which adjusts the pixel sampling current. Both of these settings are controlled with bias voltage adjustments modified by electronically changing a bank of analog switches via two 2-bit settings ({\tt vnl1sel} and {\tt vnl2sel} respectively) in the NuASIC's command register. {\tt Vnl1sel} can be adjusted over a 40 mV range while {\tt vnl2sel} modifies the vnl2 bias current as a fraction of the vnl1 bias current. Modifying these global pixel preamplifier settings allows us to counteract pixel input leakage current while minimizing the standing current at the preamplifier output. By minimizing the standing current passing through the preamplifier, the NuASIC reduces shot noise in the signal readout, which manifests as rms current that scales as a square root of the standing, or DC baseline current, and which would result in larger variance in the sampling capacitor bank charge. The best set of the {\tt vnl1sel} and {\tt vnl2sel} parameters will result in the lowest preamplifier standing current while still maintaining stable NuASIC readout. However, since the settings are applied globally the optimal parameters are a compromise between the performances of all of the individual 1024 pixels.  The {\tt vnl1sel} and {\tt vnl2sel} parameters are summarized in Table \ref{table:1} alongside the three remaining global NuASIC parameters described below.

The shaping amplifier bias current is modified by adjusting the vpl2 and vpl3 bias currents, which in turn adjusts the amplifier pulse height and settling time. These bias currents are adjusted by changing the vpl2 and vpl3 bias voltages over a 375 mV range with commandable onboard 4-bit digital-to-analog converters (DACs) ({\tt vpl2sel} and {\tt vpl3sel}) respectively. It is important for the shaping amplifier signal level to be linear with respect to input voltage and have a fast settling time, as otherwise the discriminator trigger threshold also becomes non-linear. Non-linearity hinders the choice of a global discriminator value that accurately sets the desired energy threshold. The best set of shaping amplifier command register settings -- and therefore discriminator linearity -- is thermally sensitive, and at lower temperatures ($\sim$0$^{\circ}$\,C), modified {\tt vpl2sel} and {\tt vpl3sel} values will most likely be used.

The NuASIC may additionally be run in CPM which reduces signal noise in exchange for enhanced sensitivity to leakage current. We anticipate that operating with 3~mm thick CZT produced by Redlen Technologies and a -400V bias will result in a predicted 10-20 nA of leakage current through the ASIC, which meets CPM's leakage current requirement (200~pA vs 400~pA maximum per pixel) under the assumption of uniform leakage current. Normal mode operation returns feedback current (a fraction of the output) directly to the preamplifier input. CPM instead uses a charge injection circuit to return $\sim$1/20th of the normal mode feedback current to reduce standing current at the preamplifier input  (and shot noise in the readout) while still maintaining leakage current balancing. This reduction in feedback current is performed with a capacitive divider circuit that must be periodically (1~kHz frequency) reset from a voltage reference (by disconnecting from the preamplifier input and connecting to the reference) intended to match the preamplifier input voltage. This circuit must be periodically reset or the preamplifier feedback cannot correctly balance input charge and the preamplifier baseline begins to diverge. The switching action to maintain the capacitive divider circuit causes small DC bias fluctuations at the preamplifier input which manifest as voltage ramping on the amplifier output. This 4V (actually a range of 3.92V to 4.02V) reference can have its voltage modified to reduce the DC bias fluctuations by commanding a 7-bit on-chip DAC with the {\tt 4Vsel} settings from the command register. This baseline voltage ramping and reset pattern can be seen in CPM's example preamplifier analog output in Figure \ref{f:eventexample}a. The slope of this voltage ramping trends across the NuASIC, likely as a function of pixel distance from the NuASIC ADC and readout time. While these perturbations therefore cannot be nulled entirely as any local optimization with {\tt 4Vsel} will negatively affect non-adjacent pixels, they can still be minimized.

In this work, we describe the effect of tuning the five NuASIC parameter registers listed in Table \ref{table:1} for operation in both normal mode and CPM. We compare our results against optimal values derived for the \textit{NuSTAR} mission's gain settings. In order of operation, the {\tt vnl1sel} and {\tt vnl2sel} parameters are modified together to adjust the preamplifier baseline. The combination of these 2-bit registers results in 16 potential preamplifier operation states. {\tt vpl2sel} and {\tt vpl3sel} are 4-bit registers that adjust the shaping amplifier vpl2 and vpl3 bias voltages respectively and modify the voltage profile passed to the event trigger discriminator. While operating in CPM, minimization of the preamplifier output voltage ramping is performed with the 7-bit {\tt 4Vsel} register by choosing a global voltage setting that minimizes DC bias fluctuations at the preamplifier input for the majority of NuASIC pixels.

%\newpage
%\clearpage

\section{Results of Normal Mode and CPM Testing}

\begin{figure*}[tbh!]
   \centering
   \includegraphics[width=6.5in]{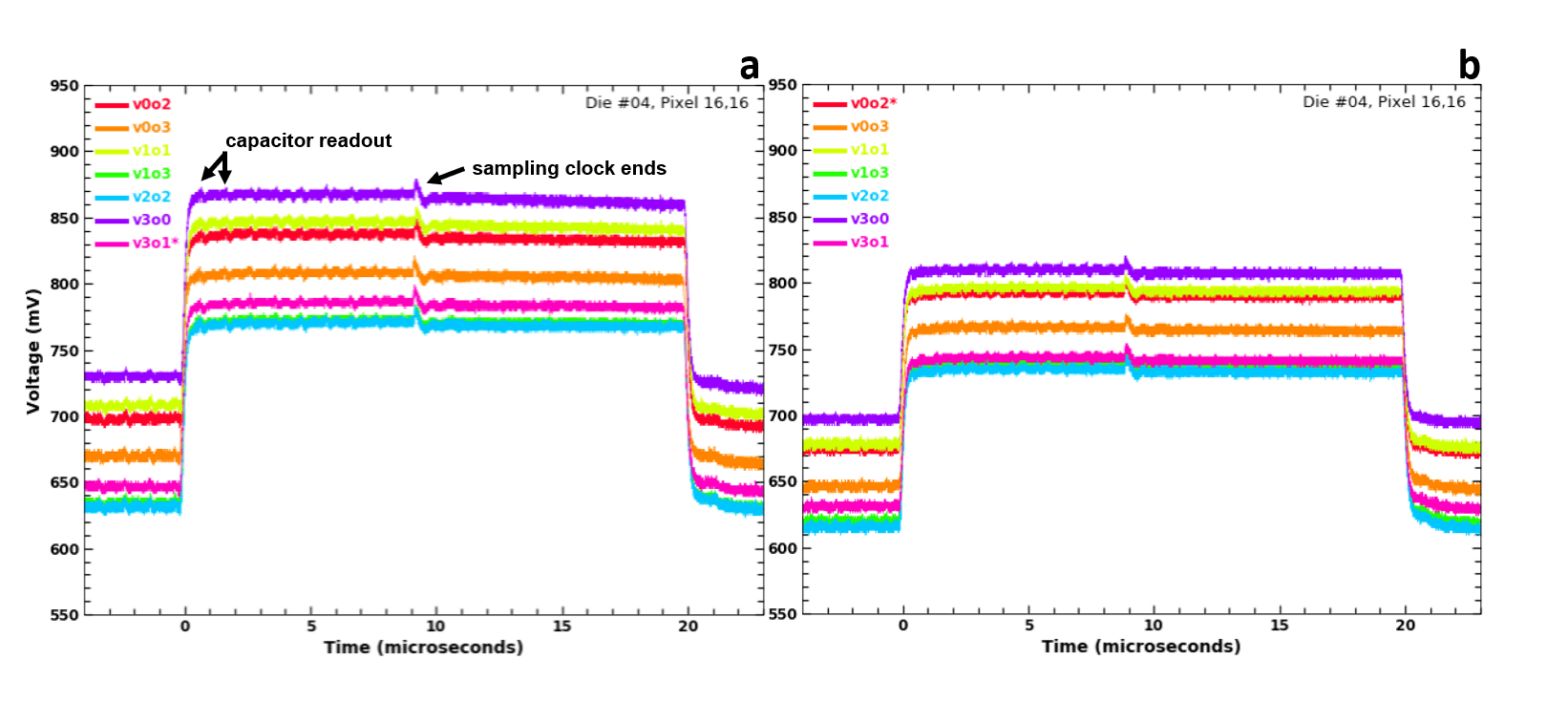}
\caption{Oscilloscope readout of NuASIC preamplifier output during pulser testing in (\textbf{\textit{a}}) normal mode and (\textbf{\textit{b}}) CPM. Modifications to the {\tt vnl1sel} and {\tt vnl2sel} parameters adjust the preamplifier standing current which in turn modifies the baseline voltage of pixel readout. Event energy corresponds to a voltage shift between baseline and event. After eight baseline and eight signal capacitors are read out (observed in bumps in the oscilloscope trace), the sampling clock stops (spike at center of signal) and post-event processing returns detector to initial state. The labels denoted by an asterisk are recommended values used by \textit{NuSTAR}. In normal mode, v3o1 achieves the lowest stable operating baseline. In CPM, a baseline is chosen (v0o2) similar to the baseline used in normal mode.}
    \label{f:preamp_traces}
\end{figure*}

Analog outputs from the preamplifier under normal mode and CPM conditions were first collected with an oscilloscope and are shown in Figure \ref{f:preamp_traces}. During testing, the {\tt vnl1sel} and {\tt vnl2sel} parameters were modified, which resulted in a shift in the preamplifier voltage baseline from approximately 630 mV to 730 mV in normal mode, shifting the floor of the signal without modifying the pulse height. Figure \ref{f:preamp_traces}a includes preamplifier traces from normal mode while Figure \ref{f:preamp_traces}b are from CPM with identical {\tt vnl1sel} and {\tt vnl2sel} settings, but resulting in slightly different baseline values. Before a signal is triggered, the sampling clock is continuously running on the baseline and can be seen as periodic 1 $\mu$s perturbations on the signal. These perturbations continue for 8 $\mu$s after the event trigger as the signal energy is recorded onto the final eight sampling capacitors. After readout this ultimately allows for a calculation of event energy by subtracting the signal from the baseline. The baseline value from the 12-bit ADC readout shifts from 524 analog-to-digital units (ADUs) to 1044 ADUs over the 630 to 730 mV range. A full record of analog baseline values for the tested parameters are listed in Table \ref{table:2} alongside keywords for their associated {\tt vnl1sel} and {\tt vnl2sel} register value combinations. At 9 $\mu$s after the event trigger, the sampling clock stops and the NuASIC enters into a locked post-trigger phase before readout. In both normal mode and CPM, {\tt vnl1sel} and {\tt vnl2sel} setting combinations that resulted in baselines lower than those shown in Figure \ref{f:preamp_traces} produced fluctuating baseline levels. These fluctuating baselines generated noise triggers and lowered detector performance by reducing signal resolution and generating additional ``dead time'' as pixel readout time increased. CPM is overall less affected by the baseline change due to the modified feedback network, and a baseline value can be chosen to approximately match the normal mode value (v0o2, the value used by \textit{NuSTAR}, or v0o3). 

\begin{figure*}[tbh!]
   \centering
   \includegraphics[width=4in]{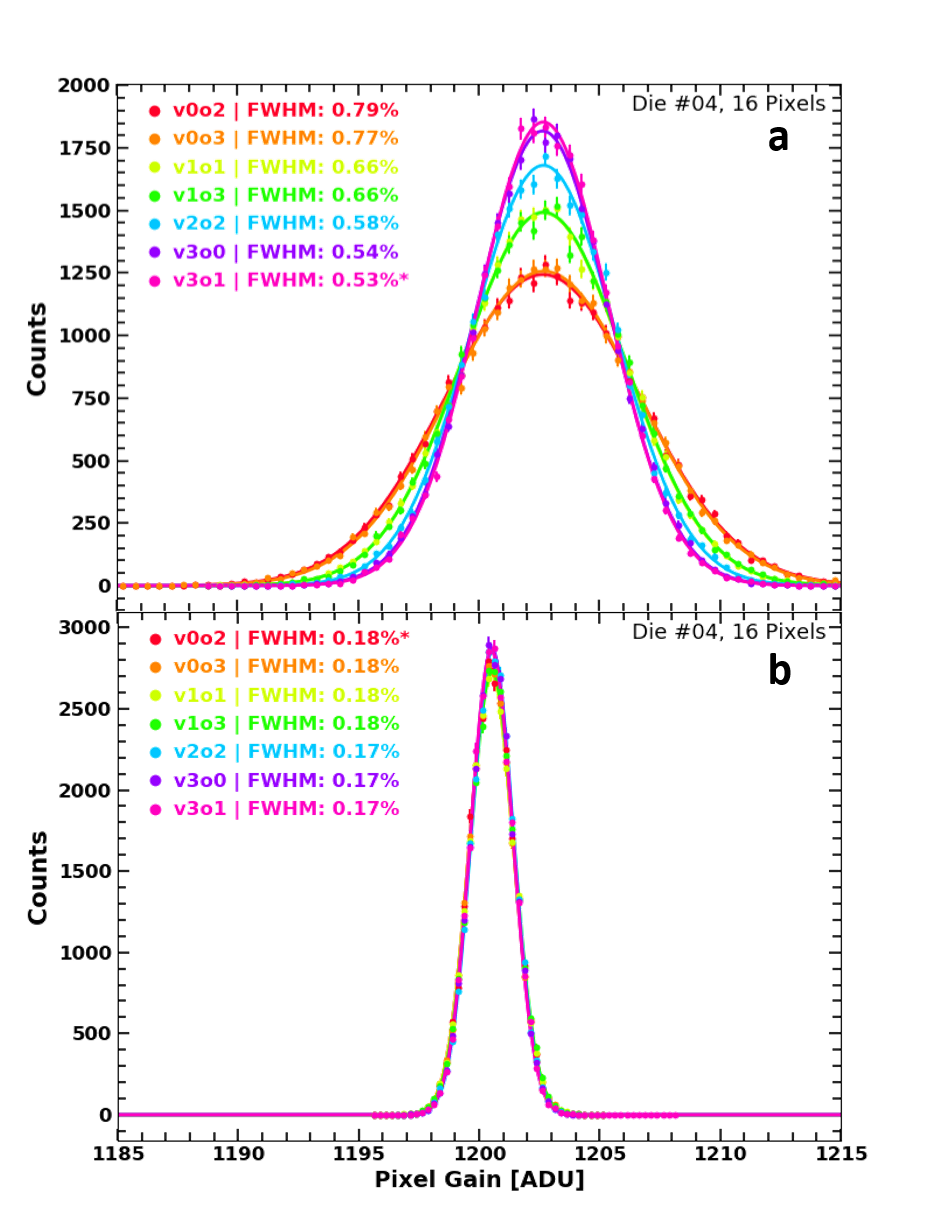}
\caption{Report of internal pulser testing of NuASIC with a variety of preamplifier settings for (\textbf{\textit{a}}) normal mode and (\textbf{\textit{b}}) CPM. The lowest stable baseline settings for normal mode is v3o1 and provides the best spectral resolution at 0.53\% FWHM (6.3 ADU). While CPM resolution is not heavily affected by preamplifier setting adjustments (ranging from 0.17-0.18\% FWHM within error), it provides a 3-fold improvement in resolution over normal mode operation. Parameter settings used by \textit{NuSTAR} are denoted with an asterisk.}
    \label{f:preamp_settings}
\end{figure*}

The results of modifying the preamplifier {\tt vnl1sel} and {\tt vnl2sel} parameters and adjusting the baseline can be observed in Figure \ref{f:preamp_settings}. We tested a patch of 16 pixels with the NuASIC's internal test pulser with various combinations of preamplifier register settings. Each pixel's gain was calibrated and the resolution of the patch was calculated from the distribution of event energies and compared across different combinations of register settings. In Figure \ref{f:preamp_settings}a, normal mode resolution is directly improved by reducing the standing current baseline of the preamplifier. The best results are achieved at \textit{NuSTAR's} normal mode default register values of {\tt vnl1sel} 3 and {\tt vnl2sel} 1 (v3o1) with a 0.53\% Full Width at Half Maximum (FWHM) resolution. The parameter with the most noticeable effect on pixel resolution is {\tt vnl1sel}, with each register step increase improving resolution by $\sim$0.1\% FWHM. The {\tt vnl2sel} parameter, while contributing to baseline shifts in Figure \ref{f:preamp_traces}, has a less noticeable effect on the pixel channel resolution. In Figure \ref{f:preamp_settings}b, CPM demonstrates an improved resolution of 0.18\% FWHM at the mode's default settings (v0o2). While operating in CPM, there was no significant resolution change from modifying {\tt vnl1sel} and {\tt vnl2sel}. This is most likely because the CPM feedback mechanism returns only a fraction ($\sim$1/20th) of the standing current at the preamplifier output.

\begin{figure*}[tbh!]
   \centering
   \includegraphics[width=6.0in]{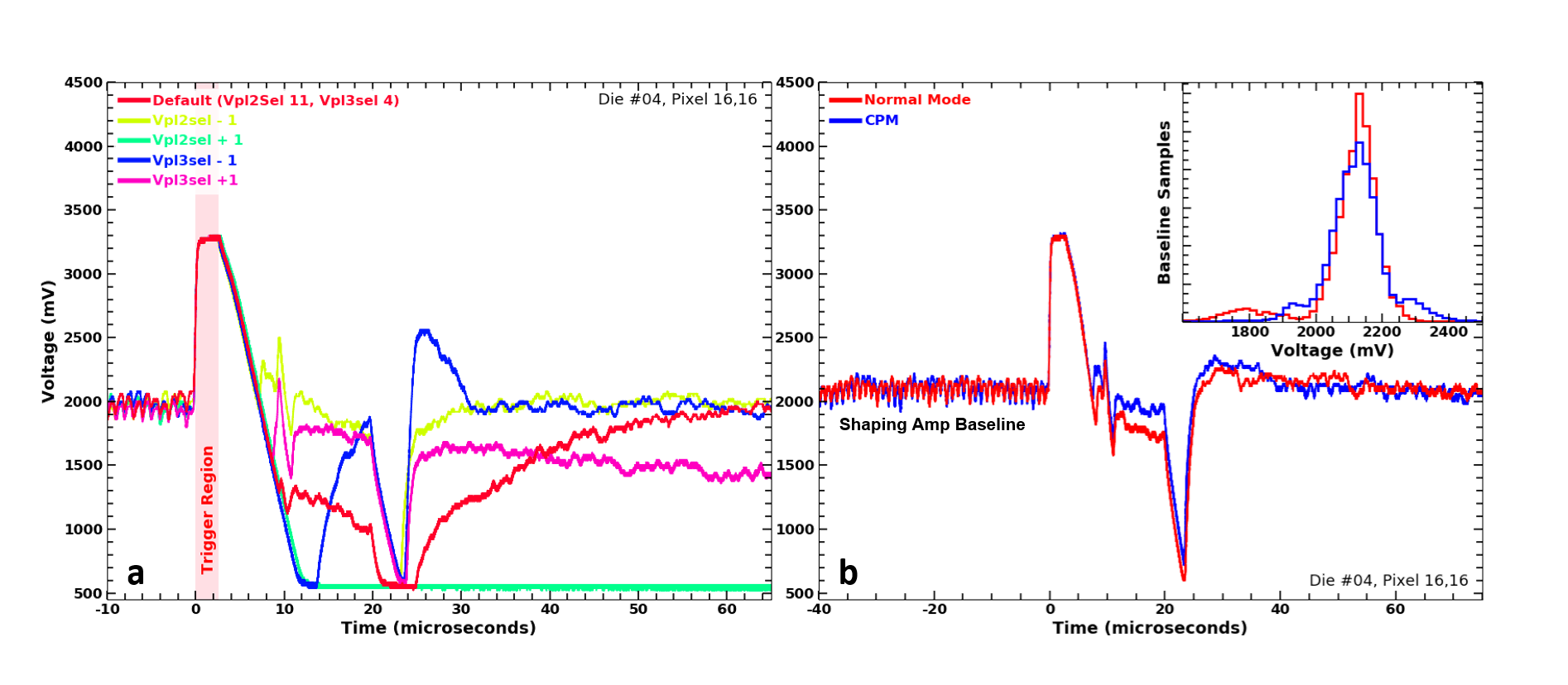}
\caption{ (\textbf{\textit{a}}) Oscilloscope readout of shaping amplifier for a variety of parameter settings around the optimal values used by \textit{NuSTAR}. Modifying the {\tt vpl2sel} and {\tt vpl3sel} register value (and vpl2 and vpl3 bias voltages, respectively) affect peak linearity and subsequent settling time. The discriminator trigger region occurs immediately once a set voltage threshold is met.  (\textbf{\textit{b}}) Shaping profiles in CPM and normal mode are nearly identical. No additional reductions of the shaping amplifier baseline noise level are obtained for a ``bare'' NuASIC operating in CPM, which would allow more sensitive event triggering. The inset provides a histogram of the baseline sample distribution and demonstrates this noise level similarity between modes. Additional improvements may be observed in systems with an attached detector.}
    \label{f:shaper}
\end{figure*}

We also explored a selection of shaping amplifier bias voltages for the NuASIC with command register settings listed in Table \ref{table:3}. By modifying the bias voltages with {\tt vpl2sel} and {\tt vpl3sel}, a variety of shaping amplifier profiles can be obtained. Records of oscilloscope-captured analog output traces from the shaping amplifier can be seen in Figure \ref{f:shaper}. The pulse height of the signal is important as the discriminator will latch the NuASIC depending on its amplitude and corresponding discriminator value. The shaping amplifier should ultimately be adjusted to have a linear response to changing signal strength, while quickly returning to a stable baseline value. Adjustments to {\tt vpl2sel} and {\tt vpl3sel} primarily modify the height and duration of the primary peak and the duration of time before the signal is returned to the baseline, and are shown in  Figure \ref{f:shaper}a. The default register settings of {\tt vpl2sel} = 11 and {\tt vpl3sel} = 4 (3.850V and 4.475V, respectively) were tested by \textit{NuSTAR} to yield a linear response across the discriminator energy range. Future work will explore the effect of modifying the shaping amplifier parameters under different thermal conditions we may expect during a HREXI flight mission. We compared analog output traces from the shaping amplifier in both normal mode and CPM, which is shown in Figure \ref{f:shaper}b. We expected that while operating in CPM the shaping amplifier baseline before the signal would be less noisy and would allow for easier separation of weak signals from the background. This reduction in baseline shaping amplifier noise was not observed. We note that without the attachment of a detector under bias, that CPM does not further improve the low energy threshold, however further testing will be required to make this determination for actual detectors systems and will be carried out in subsequent work. 

\begin{figure*}[tbh!]
   \centering
   \includegraphics[width=6.5in]{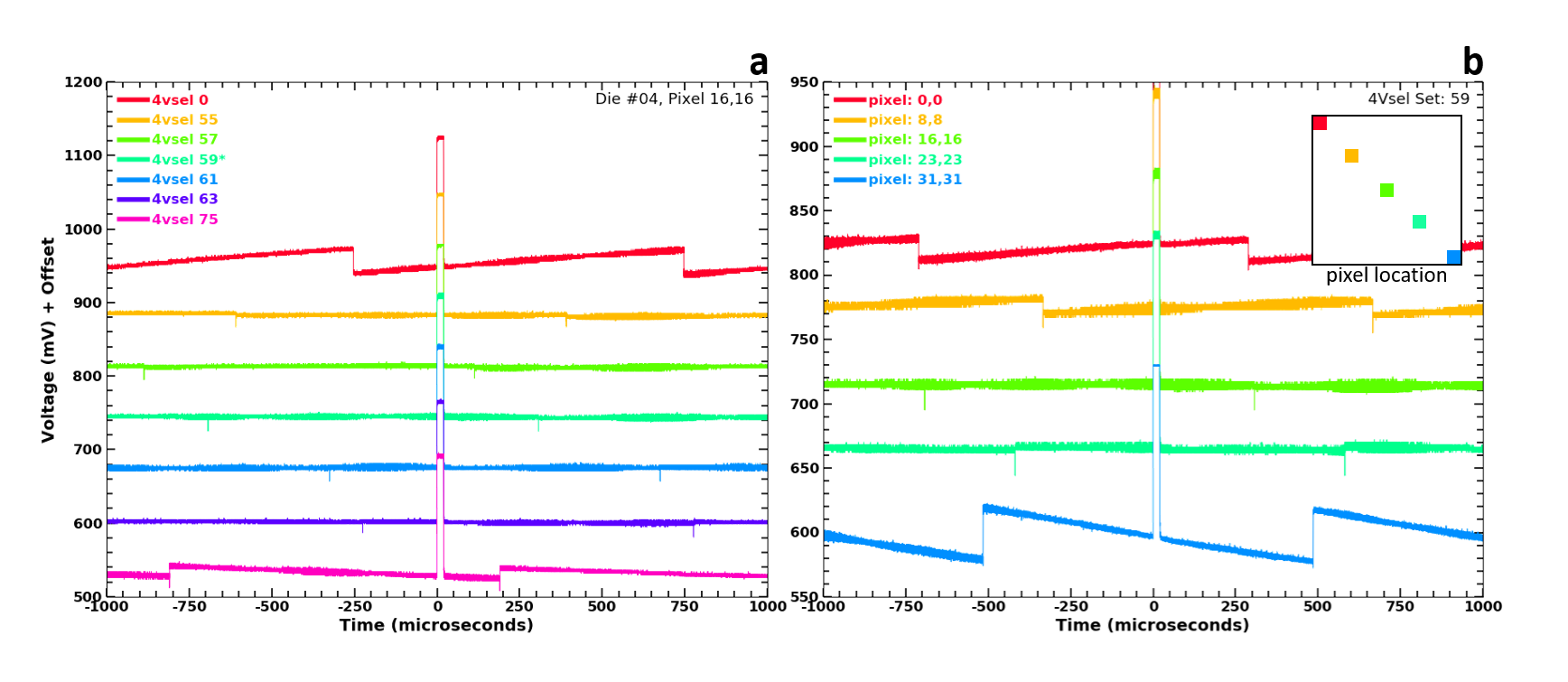}
\caption{Oscilloscope readout of NuASIC preamplifier output during pulser testing operating in CPM with a baseline offset applied for improved visibility. The oscilloscope trace duration has been chosen to show the reset of the baseline level caused by the CPM reset functionality within the NuASIC. (\textbf{\textit{a}}) A single central pixel is shown to demonstrate the impact of {\tt 4Vsel} tuning.  Modifying the {\tt 4Vsel} tuning parameter adjusts the slope of the baseline, allowing us to minimize the effects of CPM reset across the device. (\textbf{\textit{b}}) Preamplifier outputs for a fixed {\tt 4Vsel} register value are reported from five pixels across the NuASIC (pixel 1 at (0,0) to pixel 1024 at (31,31)). Pixels near the center have less voltage ramping while pixels near the edge are still affected. The insert denotes the relative position of the pixels of the same color.}
    \label{f:v4sel_adjust}
\end{figure*}

In addition to these four tuning parameters that can be explored in the NuASIC's normal mode operation, CPM introduces a feedback charge reset voltage level that can be modified by the {\tt 4Vsel} register to precisely adjust an internal $\sim$4V reference. The impact of the CPM feedback reset appears as voltage ramping in the preamplifier outputs, and can trend both positively and negatively as can be seen in Figure \ref{f:v4sel_adjust}. This figure is a longer time-scale snapshot of the analog preamplifier output that is shown in Figure \ref{f:preamp_traces}, with the pulser event signal located at the center of the oscilloscope trace. In Figure \ref{f:v4sel_adjust}a, a NuASIC pixel's oscilloscope trace is shown while operating in CPM with various {\tt 4Vsel} register settings. As the {\tt 4Vsel} register value changes from 0 to 75, the magnitude and slope of the voltage ramping effect changes, with optimal ``flat'' readout occurring between register values of 56 to 61 (4V reference voltages of 3.975 and 3.972, respectively). {\tt 4Vsel} register values greater than 75 had a tendency in the current firmware version to cause reset-induced fake event triggers due to the large positive edge of the reset signal. This will be corrected in the next firmware update, although {\tt 4Vsel} register values greater than 75 will not be typically used. Figure \ref{f:v4sel_adjust}b demonstrates how the preamplifier baseline voltage ramping effect is not flat across the device and trends from pixel (0,0) to (31,31) by sampling analog preamplifier traces from several pixel locations across the device. This trending still exists after minimizing the ramping effect on the central pixels. Although the ensemble of 1024 NuASIC pixel baseline slopes are minimized for this device at a {\tt 4Vsel} value of 59, the effect of this voltage ramping will need to be accounted for in pixels at the edges of each NuASIC.

\section{Discussion of Optimal Register Settings} \label{s:performance}

Under both normal mode and CPM operation, the preamplifier {\tt vnl1sel} and {\tt vnl2sel} parameter settings confirmed that the \textit{NuSTAR} default values (normal mode:\,v3o0; CPM:\,v0o2) continue to be optimal for HREXI's chosen gain setting and larger full-scale energy range. In normal mode, attempting to use register settings that result in a lower baseline ({\tt vnl1sel} = 0, {\tt vnl2sel} = 0/1) typically caused unstable NuASIC operation.  As a result, fluctuations that appear in the baseline generate a large number of false triggers, increasing the detector's ``dead time''. Similar effects were observed while operating in CPM, with both the lowest baselines and highest baselines ({\tt vnl1sel} = 3, {\tt vnl2sel} =3) becoming unstable. Optimal register values for the shaping amplifier have also not changed, with register values of {\tt vpl2sel} and {\tt vpl3sel} of 11 and 4 respectively for room temperature testing. Further exploration of these values with a bonded CZT detector will be carried out in future work in our Thermal Vacuum Chamber (TVAC).  This will enable us to repeat these experiments and confirm the linearity of the shaping amplifier output under realistic flight-like conditions at lower operating temperatures.

\begin{figure*}[tbh!]
   \centering
   \includegraphics[width=6.5in]{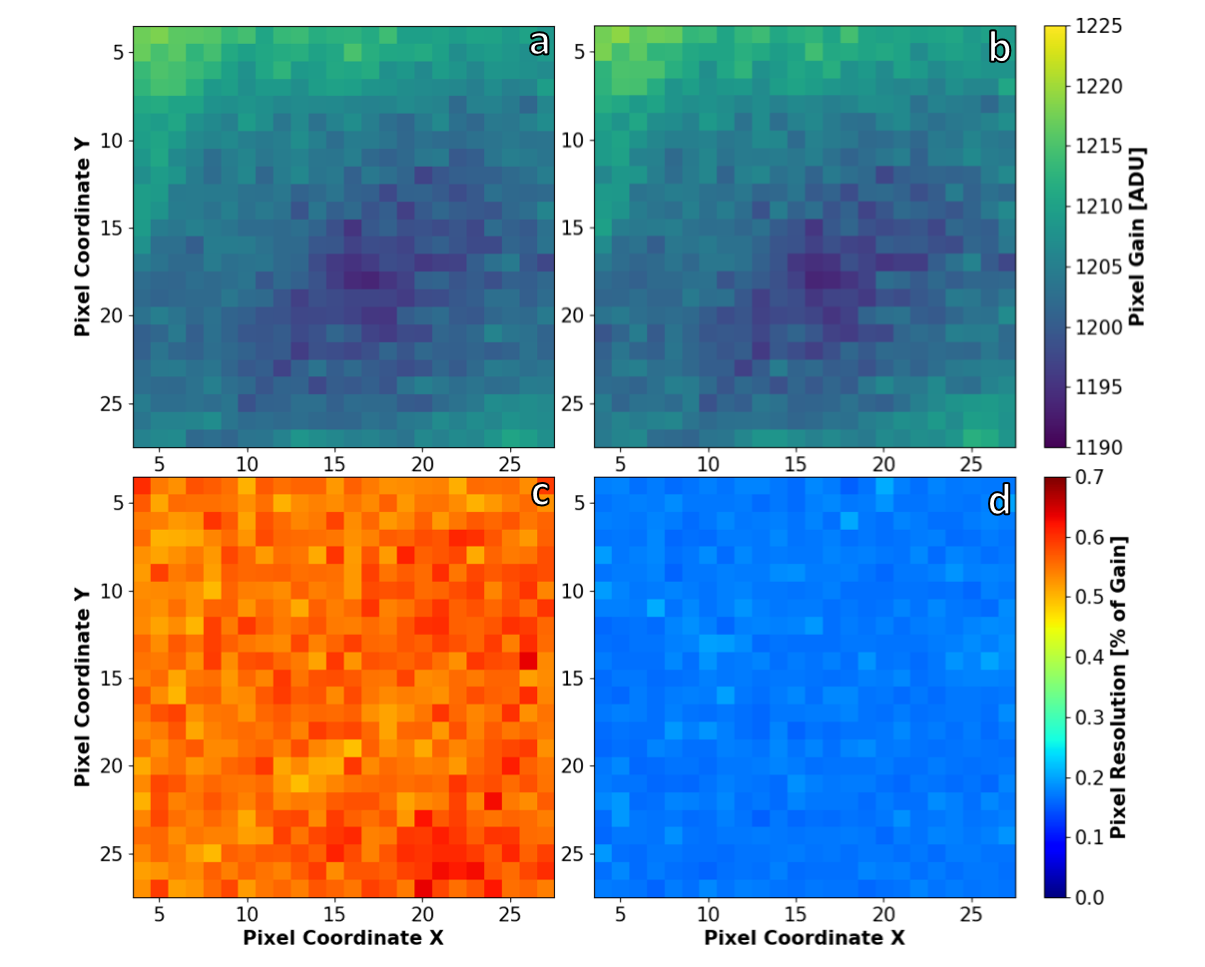}
\caption{Full ATS pixel scans of the central 576 pixels of a TSV-enabled NuSTAR ASIC comparing the normal operating mode gain (\textbf{\textit{a}}) and FWHM resolution (\textbf{\textit{c}}) with CPM gain (\textbf{\textit{b}}) and FWHM resolution (\textbf{\textit{d}}). The gain pattern across the detector is due to NuASIC pixel channel variance and routing distance from the NuASIC's 12-bit ADC. Gain variance is calibrated for the resolution maps. CPM improves the FWHM resolution of each pixel channel by a factor of $\sim$3, which will in turn improve spectral energy resolution when bonded to a detector.}
    \label{f:pulserscan}
\end{figure*}

\begin{figure*}[tbh!]
   \centering
   \includegraphics[width=6.0in]{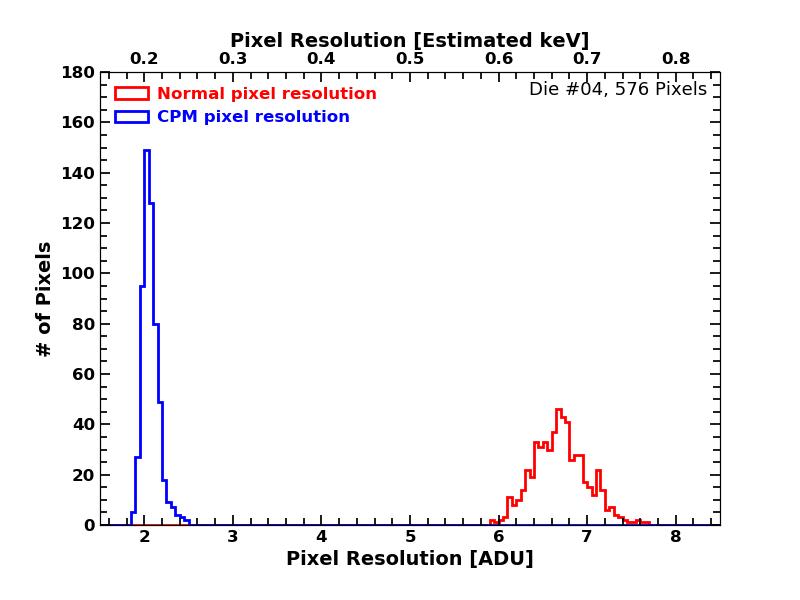}
\caption{Histogram of pixel resolutions under an internal pulser scan for normal mode operation (red) and CPM (blue) of the central 576 pixels. CPM mean pixel resolution is 2.1 ± 0.1 ADU while normal mode pixel resolution is 6.7 ± 0.3 ADU. CPM resolution improvement is strictly from the NuASIC's electronic noise and any contributing noise from the ATS test set-up, as no detector is attached. The improvement of both the median pixel resolution as well as the spread of pixel resolutions are important outcomes from CPM operation.}
    \label{f:res_histogram}
\end{figure*}

Enabling CPM with a optimally selected {\tt 4Vsel} parameter results in a 3-fold improvement in pixel channel resolution on a ``bare'' NuASIC. The mean pixel channel resolution improved from 6.7 ± 0.3  ADU ($\sim$0.7 keV) under optimal normal mode settings to 2.1 ± 0.1 ADU ($\sim$0.2 keV) when operating in CPM. The resolution and gain performance of the central 576 (24\,$\times$\,24) pixels of the NuASIC can be seen in Figure \ref{f:pulserscan}. This NuASIC pixel resolution and gain map was acquired by activating the NuASIC pixels' internal test pulsers in 4\,$\times$\,4 patches and recording the mean and variance of the pulser event energy output. The 4\,$\times$\,4 pulser patch was stepped across the NuASIC until all of the central 576 pixels were scanned, while noise-generating edge pixels external to the region under test were disabled via a commandable pixel mask. Figures \ref{f:pulserscan}a and \ref{f:pulserscan}c include the gain and resolution data while the NuASIC operates in normal mode while \ref{f:pulserscan}b and \ref{f:pulserscan}d similarly include the pixel gain and resolution data of CPM. Across the NuASIC, pixel resolution improves by a factor of 3 while pixel gain remains nearly identical, with uncalibrated pixel gain median values shifting no more than ± 3 ADU. The uneven pixel gain pattern across the NuASIC is due to the inherent variance in properties of the independent pixel channel preamplifiers and routing paths to the NuASIC 12-bit ADC. The pixel gain pattern ultimately is calibrated, resulting in the pixel resolution results shown (Figure \ref{f:pulserscan}). A histogram containing the resolution data of all pixels scanned in the pixel map is shown in Figure \ref{f:res_histogram}. The binned pixel data demonstrate that CPM improves the median NuASIC pixel resolution, as well as improves the resolution uniformity across the pixels.

\begin{figure*}[tbh!]
   \centering
   \includegraphics[width=6.0in]{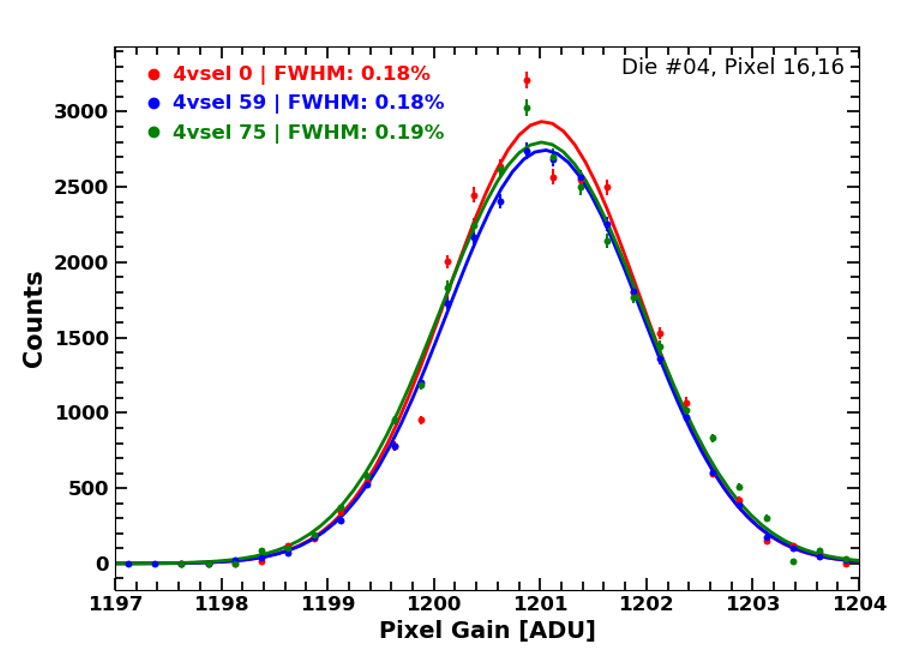}
\caption{Report of internal pulser testing of NuASIC operating in CPM with several {\tt 4Vsel} parameter settings. Even with extreme settings chosen ({\tt 4Vsel} = 0, 75) resulting in large baseline voltage ramping slopes, no change in resolution is observed under low-background and low leakage current conditions with no attached detector. Operating with {\tt 4Vsel} values above $\sim$80 resulted in unstable device operation due to a current firmware bug.}
    \label{f:cpm_4Vsel}
\end{figure*}

This improvement in resolution will ultimately be affected by the optimally chosen {\tt 4Vsel} register value. Figure \ref{f:cpm_4Vsel} includes resolution data from the pixel (16,16) data presented in Figure \ref{f:v4sel_adjust}a for the most optimal {\tt 4Vsel} value of 59, as well as values with strong positive and negative voltage ramping ({\tt 4Vsel} of 0 and 75, respectively). While testing on the ATS results in little resolution degradation for this pixel at sub-optimal {\tt 4Vsel} selections, the impact is expected to be much greater after bonding to a CZT detector and while operating under a high voltage bias. The additional leakage current and greater pixel readout rates (affecting CPM reset timing) are expected to induce additional noise in the detector, for which this {\tt 4Vsel} optimization will be important.

\begin{figure*}[tbh!]
   \centering
   \includegraphics[width=6.0in]{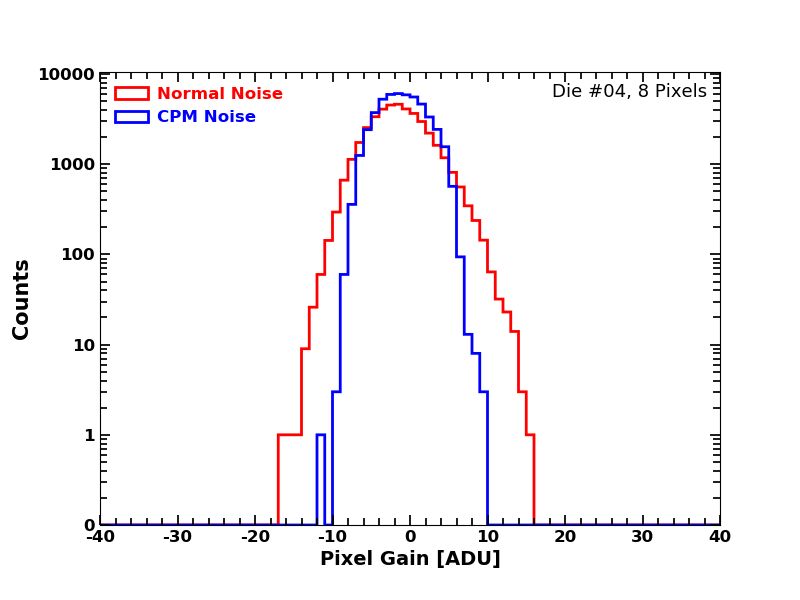}
\caption{Results of zero-point noise threshold exploration in normal mode operation (red) and CPM (blue). The distributions are the zero-point noise from noise triggers with a low discriminator ($\sim$3.5keV, register value 10 out of 255), with CPM resulting in a 6 ADU reduction in noise distribution spread over normal mode. Any noise floor below $\sim$20 ADU will allow for a high signal-to-noise detection of the minimum trigger threshold of 2 keV photons. These noise distributions may broaden with an attached detector and additional leakage current, but CPM operation is still expected to provide an improvement over normal mode operation.}
    \label{f:threshold}
\end{figure*}

Measurements of the noise distribution are performed for both optimal normal mode and CPM data. When a pixel trigger is detected on the NuASIC, we command a 3\,$\times$\,3 pixel patch readout, allowing the sampling of pixels adjacent to the triggered pixel that have no injected charge. The distribution of these noise-only samples is shown in Figure \ref{f:threshold}. We collected noise distribution data from tests performed at both a high discriminator threshold (20 keV) and a low threshold (3.5 keV). A 6 ADU reduction in noise distribution spread is observed when moving from normal mode operation to CPM while operating at a low discriminator setting (3.5 keV). The noise distribution directly affects the low energy threshold of NuASIC readout, as an ADU value of $\sim$20 corresponds to the lowest NuASIC discriminator setting at 2 keV. While the ATS is operating the NuASIC without an attached detector and the noise distribution is expected to be broader under normal detector conditions, the testing performed on the ATS confirms that operating in CPM will improve (i.e., lower) the NuASIC's low energy threshold due to electronic noise. With an attached CZT crystal, we expect our detector's noise distribution to be broader due to leakage current. In turn, we may see a larger factor of noise distribution reduction while operating in CPM which will be explored in future work.

%\newpage
%\clearpage

\section{Summary and Future Development} \label{s:futureS}

From our NuASIC testing and discussions with the \textit{NuSTAR} team, we have determined that the default parameters prescribed for the NuASIC and flown on \textit{NuSTAR} are expected to still work optimally for the modified gain settings required for the broader HREXI energy band. In our planned CPM operation, this will result in default parameter settings of {\tt vnl1sel} = 0 and {\tt vnl2sel} = 3 to set the preamplifier baseline voltage to $\sim$680\,mV. The shaping amplifier's optimal settings will also remain as the defaults at {\tt vpl2sel} = 11 and {\tt vpl3sel} = 4. \textit{NuSTAR} updated these optimal settings to as low as {\tt vpl2sel} = 9 and {\tt vpl3sel} = 4 in cooler operating temperatures ($\sim$0$^{\circ}$\,C), and we also expect to modify these settings according to the detector's flight conditions. The optimal range of {\tt 4Vsel} values for flattening preamplifier voltage ramping while operating in CPM appears to range between register values of 50 and 65, which will need to be fine-tuned for each NuASIC in the HREXI detector readout. However, by digitizing the CPM reset phase information and minimizing the voltage ramping slope in parallel for all pixels across the device with the {\tt 4Vsel} command register, we will be able to semi-automatically perform this correction.

Fabrication, testing and assembly of hundreds of CZT/NuASIC detectors will be labor intensive and every improvement to the detector integration and testing process will result in large reductions to schedule time and cost. Optimal operating parameters can only be finalized for each NuASIC after bonding with CZT, and must be performed in an environment that represents the detector's in-flight conditions. To that end, NuASIC optimization and subsequent radiation source characterization will be performed in our TVAC system at reduced temperatures to limit leakage current. This will result in characterized CZT/NuASIC detectors that are each individually optimized prior to full detector plane integration for HREXI and \textit{HSEE}.

\section{Acknowledgements}

We would like to thank Fiona A. Harrison, Jill A. Burnham, and W. Rick Cook of the California Institute of Technology along with the rest of the NuSTAR Team for developing the NuASIC and providing valuable information to facilitate its operation and testing. DPV is grateful to MA and KV for useful advice. This work was supported by NASA APRA grant NNX17AE62G. DPV is supported by the NASA FINESST Fellowship 80NSSC20K1537.

%\appendix
%\section{Appendix}
%Appendix sections are coded under \verb+\appendix+.

%\verb+\printcredits+ command is used after appendix sections to list 
%author credit taxonomy contribution roles tagged using \verb+\credit+ 
%in frontmatter.

\bibliography{references} 
%\bibliographystyle{cas-model2-names}
\bibliographystyle{spiejour}

%\bibliography{cas-refs}

\section*{Biographies}

{\bf Daniel Violette} received his PhD in astronomy and astrophysics from Harvard University in 2022, where he was supported by the Future Investigators in NASA Earth and Space Science and Technology Fellowship to further develop HREXI detector sensitivity at low energies. Currently, he is a NASA postdoctoral fellow at Goddard Space Flight Center pursuing interests in high-energy time domain astrophysics and instrumentation development.

{\bf Branden Allen} received his Ph.D.~degree in physics from U.C. Irvine in 2007 and is currently a Senior Research Scientist at Harvard University with over 20 years of experience in the development, deployment and operation of ground- and space-based telescopes for high-energy X/$\gamma$-ray astronomy and planetary science.  His current research is focused on the development and deployment of next generation detector systems and telescopes to probe high energy astrophysical phenomena and for future planetary exploration.  

{\bf Jaesub Hong} is a Senior Research Scientist at Harvard University. He has nearly 20 years of experience in development of X-ray telescopes for high energy astrophysics and planetary science.  His current focus is  the development of advanced hard X-ray detectors for next generation wide-field hard X-ray telescopes for time domain astrophysics and the miniature lightweight X-ray optics for planetary science.  He received a Ph.D.~degree in Physics from Columbia University. He has (co)authored over 40 publications. 

{\bf Hiromasa Miyasaka} is a staff scientist at California Institute of Technology. He received a Ph.D.~in Physics (2000) from Saitama University in Japan. He has over 20 years of experience in development of particles and X-ray detectors for the cosmic ray and high-energy astrophysics. Since 2006, his work has focused on CdZnTe and CdTe detectors and readout ASIC development. He is one of the primary detector scientists for the \textit{NuSTAR} mission.

{\bf Jonathan Grindlay} is the Robert Treat Paine Professor of Astronomy at Harvard. He received his BA in Physics from Dartmouth (1966) and PhD in Astrophysics from Harvard  (1971). He joined the Faculty in 1976 and  Chaired the Department in 1985-91 and 2001-03. His primary interest is black hole time variability, accretion physics, accreting black hole (both stellar and supermassive) populations and formation as measured with wide-field coded aperture imaging X-ray telescopes (ultimately full-sky) and optical/IR imaging/spectroscopy. He has over 434 refereed Journal papers.

\newpage

\section*{Figures}

\begin{table}[h!]
\centering
\begin{tabular}{| c | c | c | c | c |} 
 \hline
 Parameter & NuASIC Setting & Size (bits) & Adjustment Effect & Tuning Order \\ 
 \hline\hline
 vnl1 bias & {\tt vnl1sel} & 2 & \multirow{2}{*}{Preamplifier standing current} & \multirow{2}{*}{1}\\ 
 \cline{1-3}
 vnl2 bias & {\tt vnl2sel} & 2 & & \\
 \hline
 vpl2 bias & {\tt vpl2sel} & 4 & \multirow{2}{*}{Shaping amplifier pulse shape} & \multirow{2}{*}{2}\\
 \cline{1-3}
 vpl3 bias &{\tt vpl3sel} & 4 &  & \\
 \hline
 4V reference & {\tt 4Vsel} & 7 & Offset Preamp Voltage Ramping & 3 \\
 \hline
\end{tabular}
\bigskip
\caption{NuASIC adjustable parameters and effect of tuning on device operation.}
\label{table:1}
\end{table}

\begin{table}[h!]
\centering
\begin{tabular}{| c | c | c | c | c |} 
 \hline
 {\tt vn1sel} (bits) & {\tt vnl2sel} (bits) & Designation & Normal Baseline (mV) & CPM Baseline (mV)\\ 
 \hline\hline
 11 & 01 & v3o1 & 647 ± 2 & 630 ± 2 \\ 
 \hline
 11 & 00 & v3o0 & 731 ± 2 & 696 ± 2  \\
 \hline
 10 & 10 & v2o2 & 631 ± 2 & 615 ± 2 \\
 \hline
 01 & 11 & v1o3 & 634 ± 2 & 620 ± 2 \\
 \hline
 01 & 01 & v1o1 & 708 ± 2 & 677 ± 2 \\
 \hline
 00 & 11 & v0o3 & 669 ± 2 & 646 ± 1 \\
 \hline
 00 & 10 & v0o2 & 698 ± 2 & 674 ± 2 \\
 \hline
\end{tabular}
\bigskip
\caption{Comparison of a subset of preamplifier feedback bias settings across a range of potential baseline values. Designations denote shorthand descriptions of combinations of {\tt vnl1sel} and {\tt vnl2sel} register values used throughout the paper. The reported analog baseline values are pixel-dependent but trend similarly for each of the 1024 pixels on the device.}
\label{table:2}
\end{table}

\begin{table}[h!]
\centering
\begin{tabular}{| c | c | c | c | c |} 
 \hline
 {\tt vpl2sel} (bits) & {\tt vpl3sel} (bits) & Designation & Reset Time ($\mu$s) & Magnitude (mV) \\ 
 \hline\hline
 1011 & 0100 & Default &   60 & 3278 $\pm$ 12 \\ 
 \hline
 1010 & 0100 & {\tt vpl2sel} - 1 &   30 & 3273 $\pm$ 12 \\
 \hline
 1100 & 0100 & {\tt vpl2sel} + 1 &   	$>$100 & 3278 $\pm$ 12 \\
 \hline
 1011 & 0011 & {\tt vpl3sel} - 1 &  30 & 3276 $\pm$ 15 \\
 \hline
 1011 & 0101 & {\tt vpl3sel} + 1 &   	$>$100 & 3284 $\pm$ 8\\
 \hline
\end{tabular}
\bigskip
\caption{Comparison of a subset of shaping amplifier bias settings across values close to the suggested default ({\tt vpl2sel} = 11, {\tt vpl3sel} = 4). Combinations of {\tt vpl2sel} and {\tt vpl3sel} affect shaping amplifier pulse shape, duration, and magnitude. Small changes to the default bias settings primarily adjust pulse reset time.}
\label{table:3}
\end{table}

\end{document}